\documentclass[useAMS,usenatbib]{mn2e}
\usepackage{tabularx}
\usepackage{xspace}
\usepackage{graphicx}
\newcommand{\ignore}[1]{}

\providecommand{\ao}{}

\renewcommand{\ao}{adaptive optics (AO)\renewcommand{\ao}{AO\xspace}\renewcommand{\Ao}{AO\xspace}\xspace}
\newcommand{\Ao}{Adaptive optics (AO)\renewcommand{\ao}{AO\xspace}\renewcommand{\Ao}{AO\xspace}\xspace}
\newcommand{\wfs}{wavefront sensor (WFS)\renewcommand{\wfs}{WFS\xspace}\renewcommand{\wfss}{WFSs\xspace}\xspace}
\newcommand{\wfss}{wavefront sensors (WFSs)\renewcommand{\wfs}{WFS\xspace}\renewcommand{\wfss}{WFSs\xspace}\xspace}
\newcommand{\shwfs}{Shack-Hartmann \wfs (SHWFS)\renewcommand{\shwfs}{SHWFS\xspace}\xspace}
\newcommand{\dm}{deformable mirror (DM)\renewcommand{\dm}{DM\xspace}\renewcommand{\dms}{DMs\xspace}\renewcommand{\Dms}{DMs\xspace}\renewcommand{\Dm}{DM\xspace}\xspace}
\newcommand{\dms}{deformable mirrors (DMs)\renewcommand{\dm}{DM\xspace}\renewcommand{\dms}{DMs\xspace}\renewcommand{\Dms}{DMs\xspace}\renewcommand{\Dm}{DM\xspace}\xspace}
\newcommand{\Dms}{Deformable mirrors (DMs)\renewcommand{\dm}{DM\xspace}\renewcommand{\dms}{DMs\xspace}\renewcommand{\Dms}{DMs\xspace}\renewcommand{\Dm}{DM\xspace}\xspace}
\newcommand{\Dm}{Deformable mirror (DM)\renewcommand{\dm}{DM\xspace}\renewcommand{\dms}{DMs\xspace}\renewcommand{\Dms}{DMs\xspace}\renewcommand{\Dm}{DM\xspace}\xspace}

\newcommand{\lqg}{linear-quadratic-gaussian (LQG)\renewcommand{\lqg}{LQG\xspace}\xspace}
\newcommand{\shs}{Shack-Hartmann sensor (SHS)\renewcommand{\shs}{SHS\xspace}\renewcommand{\shss}{SHSs\xspace}\xspace}
\newcommand{\shss}{Shack-Hartmann sensors (SHSs)\renewcommand{\shs}{SHS\xspace}\renewcommand{\shss}{SHSs\xspace}\xspace}
\newcommand{\lgs}{laser guide star (LGS)\renewcommand{\lgs}{LGS\xspace}\renewcommand{\Lgs}{LGS\xspace}\renewcommand{\lgss}{LGSs\xspace}\xspace}
\newcommand{\lgss}{laser guide stars (LGSs)\renewcommand{\lgs}{LGS\xspace}\renewcommand{\Lgs}{LGS\xspace}\renewcommand{\lgss}{LGSs\xspace}\xspace}
\newcommand{\Lgs}{Laser guide star (LGS)\renewcommand{\lgs}{LGS\xspace}\renewcommand{\Lgs}{LGS\xspace}\renewcommand{\lgss}{LGSs\xspace}\xspace}

\newcommand{\Ngs}{Natural guide star (NGS)\renewcommand{\ngs}{NGS\xspace}\renewcommand{\Ngs}{NGS\xspace}\renewcommand{\ngss}{NGSs\xspace}\xspace}
\newcommand{\ngs}{natural guide star (NGS)\renewcommand{\ngs}{NGS\xspace}\renewcommand{\Ngs}{NGS\xspace}\renewcommand{\ngss}{NGSs\xspace}\xspace}
\newcommand{\ngss}{natural guide stars (NGSs)\renewcommand{\ngs}{NGS\xspace}\renewcommand{\Ngs}{NGS\xspace}\renewcommand{\ngss}{NGSs\xspace}\xspace}
\newcommand{\mems}{Micro-Electro-Mechanical Systems (MEMS)\renewcommand{\mems}{MEMS\xspace}\xspace}
\newcommand{\snr}{signal to noise ratio (SNR)\renewcommand{\snr}{SNR\xspace}\xspace}
\newcommand{\Moao}{Multi-object \ao (MOAO)\renewcommand{\moao}{MOAO\xspace}\renewcommand{\Moao}{MOAO\xspace}\xspace}
\newcommand{\moao}{multi-object \ao (MOAO)\renewcommand{\moao}{MOAO\xspace}\renewcommand{\Moao}{MOAO\xspace}\xspace}
\newcommand{\mcao}{multi-conjugate adaptive optics (MCAO)\renewcommand{\mcao}{MCAO\xspace}\xspace}
\newcommand{\ltao}{laser tomographic \ao (LTAO)\renewcommand{\ltao}{LTAO\xspace}\xspace}
\newcommand{\cpu}{central processing unit (CPU)\renewcommand{\cpu}{CPU\xspace}\renewcommand{\cpus}{CPUs\xspace}\xspace}
\newcommand{\cpus}{central processing units (CPUs)\renewcommand{\cpu}{CPU\xspace}\renewcommand{\cpus}{CPUs\xspace}\xspace}
\newcommand{\psf}{point spread function (PSF)\renewcommand{\psf}{PSF\xspace}\renewcommand{\psfs}{PSFs\xspace}\renewcommand{\Psf}{PSF\xspace}\xspace}
\newcommand{\psfs}{point spread functions (PSFs)\renewcommand{\psf}{PSF\xspace}\renewcommand{\psfs}{PSFs\xspace}\renewcommand{\Psf}{PSF\xspace}\xspace}
\newcommand{\Psf}{Point spread function (PSF)\renewcommand{\psf}{PSF\xspace}\renewcommand{\psfs}{PSFs\xspace}\renewcommand{\Psf}{PSF\xspace}\xspace}
\newcommand{\fpga}{field programmable gate array (FPGA)\renewcommand{\fpga}{FPGA\xspace}\renewcommand{\fpgas}{FPGAs\xspace}\xspace}
\newcommand{\fpgas}{field programmable gate arrays (FPGAs)\renewcommand{\fpga}{FPGA\xspace}\renewcommand{\fpgas}{FPGAs\xspace}\xspace}
\newcommand{\sor}{successive over-relaxation (SOR)\renewcommand{\sor}{SOR\xspace}\xspace}
\newcommand{\fdpcg}{Fourier domain pre-conditioned gradient (FDPCG)\renewcommand{\fdpcg}{FDPCG\xspace}\xspace}
\newcommand{\map}{maximum a-posteriori (MAP)\renewcommand{\map}{MAP\xspace}\xspace}

\newcommand{\elt}{Extremely Large Telescope (ELT)\renewcommand{\elt}{ELT\xspace}\renewcommand{\elts}{ELTs\xspace}\renewcommand{\eelt}{European ELT (E-ELT)\renewcommand{\eelt}{E-ELT\xspace}\xspace}\xspace}

\newcommand{\elts}{Extremely Large Telescopes (ELTs)\renewcommand{\elt}{ELT\xspace}\renewcommand{\elts}{ELTs\xspace}\renewcommand{\eelt}{European ELT (E-ELT)\renewcommand{\eelt}{E-ELT\xspace}\xspace}\xspace}

\newcommand{\eelt}{European Extremely Large Telescope (E-ELT)\renewcommand{\eelt}{E-ELT\xspace}\renewcommand{\elt}{ELT\xspace}\renewcommand{\elts}{ELTs\xspace}\xspace}

\newcommand{\dugall}{Durham University generalised adaptive optics laser laboratory (DUGALL)\renewcommand{\dugall}{DUGALL\xspace}\xspace}
\newcommand{\fwhm}{full-width at half-maximum (FWHM)\renewcommand{\fwhm}{FWHM\xspace}\xspace}
\newcommand{\wht}{William Herschel Telescope (WHT)\renewcommand{\wht}{WHT\xspace}\xspace}
\newcommand{\emccd}{electron multiplying CCD (EMCCD)\renewcommand{\emccd}{EMCCD\xspace}\renewcommand{\emccds}{EMCCDs\xspace}\xspace}
\newcommand{\emccds}{electron multiplying CCDs (EMCCDs)\renewcommand{\emccd}{EMCCD\xspace}\renewcommand{\emccds}{EMCCDs\xspace}\xspace}
\newcommand{\dasp}{Durham \ao simulation platform (DASP)\renewcommand{\dasp}{DASP\xspace}\renewcommand{\thedasp}{DASP\xspace}\renewcommand{\Thedasp}{DASP\xspace}\renewcommand{\daspcite}{DASP\xspace}\renewcommand{\daspcite}{DASP\xspace}\renewcommand{\thedaspcite}{DASP\xspace}\xspace}
\newcommand{\daspcite}{Durham \ao simulation platform \citep[DASP,]{basden5}\renewcommand{\dasp}{DASP\xspace}\renewcommand{\thedasp}{DASP\xspace}\renewcommand{\Thedasp}{DASP\xspace}\renewcommand{\daspcite}{DASP\xspace}\renewcommand{\thedaspcite}{DASP\xspace}\xspace}
\newcommand{\thedaspcite}{the Durham \ao simulation platform \citep[DASP,]{basden5}\renewcommand{\dasp}{DASP\xspace}\renewcommand{\thedasp}{DASP\xspace}\renewcommand{\Thedasp}{DASP\xspace}\renewcommand{\daspcite}{DASP\xspace}\renewcommand{\thedaspcite}{DASP\xspace}\xspace}

\newcommand{\thedasp}{the Durham \ao simulation platform (DASP)\renewcommand{\dasp}{DASP\xspace}\renewcommand{\thedasp}{DASP\xspace}\renewcommand{\Thedasp}{DASP\xspace}\renewcommand{\daspcite}{DASP\xspace}\renewcommand{\thedaspcite}{DASP\xspace}\xspace}
\newcommand{\Thedasp}{The Durham \ao simulation platform (DASP)\renewcommand{\dasp}{DASP\xspace}\renewcommand{\thedasp}{DASP\xspace}\renewcommand{\Thedasp}{DASP\xspace}\renewcommand{\daspcite}{DASP\xspace}\renewcommand{\thedaspcite}{DASP\xspace}\xspace}
\newcommand{\mpi}{Message Passing Interface (MPI)\renewcommand{\mpi}{MPI\xspace}\xspace}
\newcommand{\smp}{symmetric multi-processing (SMP)\renewcommand{\smp}{SMP\xspace}\xspace}
\newcommand{\svd}{singular value decomposition (SVD)\renewcommand{\svd}{SVD\xspace}\xspace}
\newcommand{\gpu}{graphics processing unit (GPU)\renewcommand{\gpu}{GPU\xspace}\renewcommand{\gpus}{GPUs\xspace}\xspace}
\newcommand{\gpus}{graphics processing units (GPUs)\renewcommand{\gpu}{GPU\xspace}\renewcommand{\gpus}{GPUs\xspace}\xspace}
\newcommand{\fft}{fast Fourier transform (FFT)\renewcommand{\fft}{FFT\xspace}\xspace}
\newcommand{\ifu}{integral field unit (IFU)\renewcommand{\ifu}{IFU\xspace}\xspace}
\newcommand{\darc}{the Durham \ao real-time controller (DARC)\renewcommand{\darc}{DARC\xspace}\renewcommand{\Darc}{DARC\xspace}\xspace}
\newcommand{\Darc}{The Durham \ao real-time controller (DARC)\renewcommand{\darc}{DARC\xspace}\renewcommand{\Darc}{DARC\xspace}\xspace}
\newcommand{\darccite}{the Durham \ao real-time controller \citep[DARC,]{basden9}\renewcommand{\darc}{DARC\xspace}\renewcommand{\Darc}{DARC\xspace}\renewcommand{\darccite}{DARC\xspace}\xspace}
\newcommand{\cots}{commercial off-the-shelf (COTS)\renewcommand{\cots}{COTS\xspace}\xspace}
\newcommand{\rtcp}{real-time control pipeline (RTCP)\renewcommand{\rtcp}{RTCP\xspace}\xspace}
\newcommand{\rms}{root-mean-square (RMS)\renewcommand{\rms}{RMS\xspace}\xspace}
\newcommand{\sFPDP}{serial Front Panel Data Port (sFPDP)\renewcommand{\sFPDP}{sFPDP\xspace}\xspace}
\newcommand{\wpu}{wavefront processing unit (WPU)\renewcommand{\wpu}{WPU\xspace}\xspace}

\newcommand{\rtcs}{real-time control system (RTCS)\renewcommand{\rtcs}{RTCS\xspace}\renewcommand{\rtcss}{RTCSs\xspace}\xspace}
\newcommand{\rtcss}{real-time control systems (RTCSs)\renewcommand{\rtcs}{RTCS\xspace}\renewcommand{\rtcss}{RTCSs\xspace}\xspace}
\newcommand{\eso}{European Southern Observatory (ESO)\renewcommand{\eso}{ESO\xspace}\renewcommand{\theeso}{ESO\xspace}\xspace}
\newcommand{\theeso}{\renewcommand{\theeso}{ESO\xspace}the \eso}
\newcommand{\scao}{single conjugate \ao (SCAO)\renewcommand{\scao}{SCAO\xspace}\renewcommand{\Scao}{SCAO\xspace}\xspace}
\newcommand{\Scao}{Single conjugate \ao (SCAO)\renewcommand{\scao}{SCAO\xspace}\renewcommand{\Scao}{SCAO\xspace}\xspace}
\newcommand{\glao}{ground layer \ao (GLAO)\renewcommand{\glao}{GLAO\xspace}\xspace}
\newcommand{\eagle}{ELT Adaptive optics for GaLaxy Evolution (EAGLE)\renewcommand{\eagle}{EAGLE\xspace}\xspace}
\newcommand{\maory}{multi-conjugate \ao relay for the \eelt (MAORY)\renewcommand{\maory}{MAORY\xspace}\xspace}
\newcommand{\muse}{Multi Unit Spectroscopic Explorer (MUSE)\renewcommand{\muse}{MUSE\xspace}\xspace}
\newcommand{\vlt}{Very Large Telescope (VLT)\renewcommand{\vlt}{VLT\xspace}\xspace}

\newcommand{\eapd}{electron avalanche photodiode\renewcommand{\eapd}{eAPD\xspace}\xspace}
\newcommand{\tmt}{Thirty Metre Telescope (TMT)\renewcommand{\tmt}{TMT\xspace}\xspace}
\newcommand{\lbt}{Large Binocular Telescope (LBT)\renewcommand{\lbt}{LBT\xspace}\xspace}
\newcommand{\xao}{eXtreme \ao (XAO)\renewcommand{\xao}{XAO\xspace}\xspace}

\newcommand{\vla}{Very Large Array (VLA)\renewcommand{\vla}{VLA\xspace}\xspace}
\newcommand{\jwst}{{\em James Webb Space Telescope} \citep[JWST,][]{jwst}\renewcommand{\jwst}{{\em JWST}\xspace}\xspace}
\newcommand{\hst}{{\em Hubble Space Telescope (HST)}\renewcommand{\hst}{{\em HST}\xspace}\xspace}
\newcommand{\ifss}{integral-field spectrographs (IFSs)\renewcommand{\ifss}{IFSs\xspace}\renewcommand{\ifs}{IFS\xspace}\xspace}
\newcommand{\ifs}{integral-field spectrograph (IFS)\renewcommand{\ifss}{IFSs\xspace}\renewcommand{\ifs}{IFS\xspace}\xspace}
\newcommand{\ifus}{integral field units (IFUs)\renewcommand{\ifus}{IFUs\xspace}\xspace}
\newcommand{\mos}{multi-object spectrograph (MOS)\renewcommand{\mos}{MOS\xspace}\xspace}
\newcommand{\goodss}{Great Observatories Origins Deep Survey (GOODS)-S\renewcommand{\goodss}{GOODS-S\xspace}\xspace}
\newcommand{\goods}{Great Observatories Origins Deep Survey (GOODS)\renewcommand{\goods}{GOODS\xspace}\xspace}
\newcommand{\cmos}{complimentary metal-oxide semiconductor (CMOS)\renewcommand{\cmos}{CMOS\xspace}\xspace}
\newcommand{\scmos}{scientific CMOS (sCMOS)\renewcommand{\scmos}{sCMOS\xspace}\xspace}
\newcommand{\aof}{Adaptive Optics Facility (AOF)\renewcommand{\aof}{AOF\xspace}\xspace}
\newcommand{\dsp}{digital signal processor (DSP)\renewcommand{\dsp}{DSP\xspace}\renewcommand{\dsps}{DSPs\xspace}\xspace}
\newcommand{\dsps}{digital signal processors (DSPs)\renewcommand{\dsp}{DSP\xspace}\renewcommand{\dsps}{DSPs\xspace}\xspace}
\newcommand{\capi}{Coherent Accelerator Processor Interface (CAPI)\renewcommand{\capi}{CAPI\xspace}\xspace}
\newcommand{\qe}{quantum efficiency (QE)\renewcommand{\qe}{QE\xspace}\xspace}
\newcommand{\numa}{non-uniform memory access (NUMA)\renewcommand{\numa}{NUMA\xspace}\xspace}
\newcommand{\uav}{unmanned aerial vehicle (UAV)\renewcommand{\uav}{UAV\xspace}\renewcommand{\uavs}{UAVs\xspace}\renewcommand{\Uav}{UAV\xspace}\xspace}
\newcommand{\uavs}{unmanned aerial vehicles (UAVs)\renewcommand{\uav}{UAV\xspace}\renewcommand{\uavs}{UAVs\xspace}\renewcommand{\Uav}{UAV\xspace}\xspace}
\newcommand{\Uav}{Unmanned aerial vehicle (UAV)\renewcommand{\uav}{UAV\xspace}\renewcommand{\Uav}{UAV\xspace}\renewcommand{\uavs}{UAVs\xspace}\xspace}

\newcommand{\ncpa}{non-common path aberration (NCPA)\renewcommand{\ncpa}{NCPA\xspace}\renewcommand{\ncpas}{NCPAs\xspace}\xspace}
\newcommand{\ncpas}{non-common path aberrations (NCPA)\renewcommand{\ncpa}{NCPA\xspace}\renewcommand{\ncpas}{NCPAs\xspace}\xspace}
\newcommand{\sdk}{software developers kit (SDK)\renewcommand{\sdk}{SDK\xspace}\renewcommand{\sdks}{SDKs\xspace}\xspace}
\newcommand{\sdks}{software developers kits (SDKs)\renewcommand{\sdk}{SDK\xspace}\renewcommand{\sdks}{SDKs\xspace}\xspace}
\newcommand{\dac}{digital to analogue converter (DAC)\renewcommand{\dac}{DAC\xspace}\xspace}
\newcommand{\nda}{non-disclosure agreement (NDA)\renewcommand{\nda}{NDA\xspace}\xspace}
\newcommand{\polc}{pseudo-open-loop control (POLC)\renewcommand{\polc}{POLC\xspace}\xspace}
\newcommand{\udp}{User Datagram Protocol (UDP)\renewcommand{\udp}{UDP\xspace}\xspace}
\newcommand{\ags}{artificial guide star (AGS)\renewcommand{\ags}{AGS\xspace}\xspace}
\newcommand{\est}{European Solar Telescope (EST)\renewcommand{\est}{EST\xspace}\xspace}
\newcommand{\lot}{Large Optical Telescope (LOT)\renewcommand{\lot}{LOT\xspace}\xspace}
\newcommand{\gtc}{Gran Telescopio Canarias (GTC)\renewcommand{\gtc}{GTC\xspace}\xspace}
\newcommand{\cta}{Cherenkov Telescope Array (CTA)\renewcommand{\cta}{CTA\xspace}\xspace}
\newcommand{\rtk}{Real-time Kinematic (RTK)\renewcommand{\rtk}{RTK\xspace}\xspace}
\newcommand{\gnss}{global navigation satellite systems (GNSS)\renewcommand{\gnss}{GNSS\xspace}\xspace}
\newcommand{\race}{Rapid Automatic Cascode Exchange (RACE)\renewcommand{\race}{RACE\xspace}\xspace}
\newcommand{\tvm}{total variation minimisation (TVM)\renewcommand{\tvm}{TVM\xspace}\xspace}

\usepackage{amssymb}

\title[Artificial guide stars using UAVs]{Artificial guide stars for
  adaptive optics using unmanned aerial vehicles}

\author[A.\ G.\ Basden et al.]{A.\ G.\ Basden$^{1}$\thanks{E-mail:
    a.g.basden@durham.ac.uk (AGB)}, Anthony M.\ Brown$^{1}$,  P.\ M.\ Chadwick$^{1}$,
    P.\ Clark$^{1}$ and
  R.\ Massey$^{1}$
\\
$^{1}$Centre for Advanced Instrumentation, Department of Physics, South Road, Durham, DH1 3LE,
UK
}
\begin{document}
\maketitle

\begin{abstract}
  Astronomical adaptive optics systems are used to increase effective
  telescope resolution.  However, they cannot be used to observe the
  whole sky since one or more natural guide stars of sufficient
  brightness must be found within the telescope field of view for the
  AO system to work.  Even when laser guide stars are used, natural
  guide stars are still required to provide a constant position
  reference.  Here, we introduce a technique to overcome this problem
  by using rotary unmanned aerial vehicles (UAVs) as a platform from
  which to produce artificial guide stars.  We describe the concept,
  which relies on the UAV being able to measure its precise relative
  position.  We investigate the adaptive optics performance
  improvements that can be achieved, which in the cases presented here
  can improve the Strehl ratio by a factor of at least 2 for a 8~m
  class telescope.  We also
  discuss improvements to this technique, which is relevant to both
  astronomical and solar adaptive optics systems.
\end{abstract}


\section{Introduction}

\Uav technology has been developing rapidly in recent years,
particularly for rotary vehicles (e.g.\ quadcopters, hexacopters and
octocopters).  In particular, this has been driven by advances in
battery performance, carbon fibre technology, and microcontroller
advances.  These advances are likely to continue in the foreseeable
future, and so here we consider a potential use for \uav technology
within astronomy, a field where adoption is relatively new.
\citet{2016SPIE.9912E..10B} and \citet{2015PASP..127.1131C} describe
\uav based systems that are suitable for calibration of telescopes.
Early work by the Pierre Auger Observatory Cosmic Ray Detector took
advantage of these advances to characterise the individual pixel
sensitivities of their fluorescence telescopes \citep{bauml}. In
particular, an omnidirectional light source was mounted on an
octocopter \uav, and positioned 500--1000~m above the telescopes, and
individual pixels of these fluorescence telescopes were
illuminated. Using the same calibration payload, the sensitivity of
other cosmic ray fluorescence detectors were cross-calibrated
\citep{matthews, hayashi}.  Building on this work, the feasibility of
using \uavs to calibrate the Cherenkov Telescope Array has also been
investigated, including detailed flight characterisation of the \uavs
to quantify their contribution to the overall systematic uncertainty
of the technique \citep{brown1,brown2}. Other examples of using UAVs
to calibrate telescope instrumentation include the characterisation of
the far-field beam map of three CROME microwave reflector antennas
\citep{felix} and the characterisation of the far-field beam map of a
single radio dish \citep{chang}.  Fixed wing \uavs have also been
proposed as a method of returning astronomical data from a long
duration super-pressure balloon airborne telescope, SuperBIT
\citep{superbit}.  Here, we explore the potential that this \uav
technology has for the field of \ao, including both solar \ao and
astronomical \ao, using \uavs to provide artificial guide stars.

The sky coverage of astronomical \ao systems \citep{adaptiveoptics} is
limited by the availability of guide stars with sufficient flux close
to the astronomical source of interest, even for wide field-of-view
\elt instruments \citep{basden17}.  Even \lgs \ao
systems \citep{laserguidestar} suffer from this since at least one
\ngs is required to compensate for the \lgs position uncertainty; the
true position of the \lgs is unknown due to atmospheric turbulence
encountered during upward propagation of the laser beam.

For solar \ao systems, high spatial resolution observation of the
solar limb is difficult, since there are no guide star references with
sufficient brightness to be used with a conventional wavefront sensor \citep{2015SoPh..290.1871T}.

\begin{figure}
  \includegraphics[width=\linewidth]{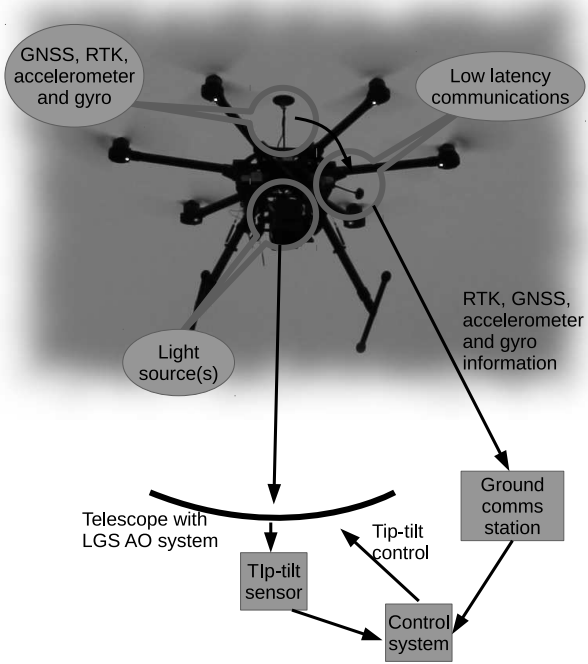}
\caption{A figure showing the UAV artificial guide star concept: a UAV
  platform is used to provide support for a light source, low latency
  communications, and instrumentation with which precise
  position knowledge can be computed (on the ground).  The UAV is
  observed using a tip-tilt sensor at the telescope, and the raw
  instrumentation data is sent via a low latency communication link to
a ground station.  This information is combined with the tip-tilt
signal so that actual atmospheric tip-tilt can be computed.  Acronym
GNSS expands to Global Navigation Satellite System, and RTK expands to
Real-Time Kinematic.}
\label{fig:uav}
\end{figure}

Here we present a solution to these problems by introducing a new form
of \ags.  For astronomical \ao systems, this \ags will provide an
absolute position (tip-tilt) signal for a \lgs \ao system, thereby
negating the requirement for \ngss entirely.  For solar \ao systems,
this \ags will provide a high order wavefront sensor target, though
without providing low order tip-tilt information.  Our proposed
techniques use an artificial light source mounted on a \uav platform,
as shown schematically in Fig.~\ref{fig:uav}.  There are two
considerations that are key to this concept.  Firstly, the relative
position stability of the UAV platform must be sufficient to enable
the light source to be maintained within the field of view of a
ground-based \wfs.  Secondly, for the astronomical case, the precise
relative position of the light source must be known instantaneously,
by some means other than optical measurement from the ground.

The concept that we are proposing will become increasingly feasible as
\uav performance increases.  Therefore we are attempting to outline a
potential future technology for \ao, rather than being restricted by
currently available \uav models.  For example, recent developments
with hydrogen fuel cells has led to bespoke \uavs with significantly
increased flight times and maximum altitudes.

\subsection{Tip-tilt correction for astronomical AO}
The concept that we introduce here sees the \uav hovering at a
significant altitude (e.g.\ 1~km or more) above a ground-based
telescope, situated along its current line of sight, tracking the
telescope motion.  The necessary information to compute the precise
relative instantaneous position of the \uav (accelerometer, \rtk and
gyroscope measurements) will be relayed to the ground using a low
latency wireless communication link, and the
relative position of a light source mounted on the \uav as seen by the
telescope will be measured by either a tip-tilt wavefront sensor, e.g.
a single sub-aperture Shack-Hartmann \wfs (for the astronomical case),
or a high order wavefront sensor (for the solar case, and possibly the
astronomical case).  By combining this optical position measurement
with the \uav derived instantaneous position estimation, the incident
wavefront tilt can be estimated and hence corrected using an active
mirror component, for example a \dm.

At good observatory sites typical astronomical seeing is around
0.7~arcseconds.  Any tip-tilt correction must lead to an improvement
over natural seeing.  In this paper, we consider a goal of
0.1~arcsecond precision in tip-tilt correction as a baseline, though
we note that this is not an intrinsic limitation of the method
described here, but one which would give a definite image quality
improvement.  For an object (i.e.\ the \uav) placed at 1~km from a
telescope, a 0.1~arcsecond resolution corresponds to a lateral
displacement of about 0.5~mm.  At least 1~km altitude is necessary for
the \uav in order for it to be (a) in the telescope's far-field limit
and (b) above most of the atmospheric turbulence.  Our baseline
performance therefore requires a relative instantaneous lateral
position knowledge of better than 1~mm for a \uav at greater than 1~km
distance.  We note that as \uav height increases, a less stringent
position knowledge is required to meet the same angular accuracy, or
alternatively, better tip-tilt estimation can be obtained.  In
\S\ref{sect:uavpos}, we discuss how instantaneous relative \uav
positional information can be obtained.  We note that the far-field
approximation may not be valid if the \uav it stationed close to
turbulent layers, and we investigate different propagation models
in \S\ref{sect:fresnel}.  However, to avoid this problem, the \uav can be operated at
heights that do not correspond to turbulent layers.

Depending on \uav flight stability, it may not be possible to maintain
the \uav source position within the \wfs field of view.  In this case,
an array of light sources can be used to ensure that at least one is
within the \wfs field of view (typically 2--10~arcseconds), with the
source closest to the centre of the field then being identified and
used by the \ao system.  These sources can be tracked in the \ao
system software so that which source is currently within the field of
view is known, and when switching between sources (i.e.\ when the \uav
position has changed significantly), the relevant slope offset
corresponding to the known change in source position can be subtracted
from the wavefront slope measurement.  It should be noted that
operation of an \ags at low altitudes will mean that only low level
turbulence tip-tilt can be measured.  We discuss this in section
\S\ref{sect:lowlevelturb}.  A key feature of a \uav \ags is the
ability to rapidly reach the desired sky position.  For current \uav
technology a 1~km altitude can be reached in 2--5 minutes, depending
on payload, telescope zenith angle and \uav model.

\subsection{Wavefront measurement for solar AO}
For solar limb studies using solar \ao systems, a high order wavefront
sensor reference is required.  The \uav will therefore track the solar
motion at a distance from the telescope of at least 1~km, maintaining
a position along the line of sight to the solar limb.  For morning
or evening observations, when the sun is low in the sky, this distance
could be significantly larger, with the UAV launched some distance
from the telescope.  We note, of course, that this would not be an
attractive option at solar observatories on isolated mountain tops
(e.g.\ Hawaii or La Palma), since a \uav launched far from the
observatory would then have to climb a significant height just to
reach the observatory altitude.  As long as the \uav source position
is maintained within the wavefront sensor field of view, and within an
isoplanatic patch size (typically 1-10~arcseconds depending on
conditions and observation wavelength), the \uav position (from which
tip-tilt information is derived) is unimportant, since faint solar
structures can be used to obtain the tip-tilt signal, using the whole
telescope aperture.  Instead, high order wavefront information will be
obtained, and used to reconstruct the high order wavefront.  The
tip-tilt (position) information from the \uav will be discarded, being
instead retrieved from a global image of the limb structure.
Additionally, in the case where science image acquisition is fast
enough (as can sometimes be the case for solar \ao), lucky imaging
techniques \citep{2006A&A...446..739L,basden16} can be used to shift
the images thereby removing the tip-tilt errors.

\begin{figure}
  \includegraphics[width=\linewidth]{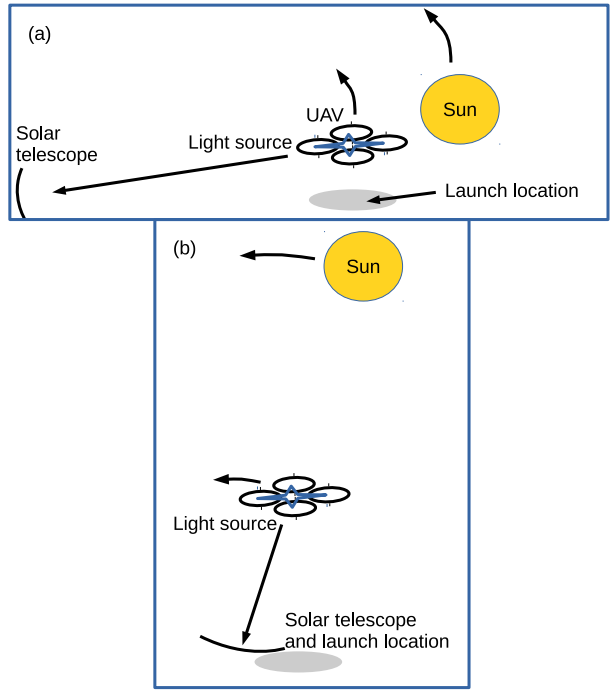}
  \caption{A figure showing the solar UAV technique.  The UAV tracks
    the solar motion as seen by the telescope, and the light source is
    viewed by a wavefront sensor at the telescope, allowing high order
    wavefront information to be retrieved.  The distance to the UAV
    can be greater when the sun is lower in the sky, due to the
    reduced altitude that the UAV has to reach.  (a) shows the case
    when the sun is low, with the UAV launched far (e.g.\ several km) from the
    telescope. (b) shows the case when the sun is higher in the sky,
    with the UAV launched from close to the telescope.}
  \label{fig:solar}
\end{figure}

The wavefront sensor will view a light source on the UAV and this
information will be used to reconstruct the incident wavefront.  The
tip-tilt information will be discarded.  We note that this is contrary
to the astronomical case, where only the low order information is
required.  Hence, the solar implementation will be technologically
less demanding, since precise position knowledge is not
necessary.  Fig.~\ref{fig:solar} demonstrates this concept, and it
should be noted that when lower in the sky, the \uav can be launched further
from the telescope, thus
reducing focal anisoplanatism, assuming some constant maximum altitude
can be reached by the \uav.  Although the sun is bright, a narrow
bandwidth light source can be used so that the \uav signal is above
that of the solar background.

\subsection{Overview of current rotary UAV technology}
In recent years, the popularity of rotary \uav technology (which we
hereafter refer to as simply \uav technology since fixed wing systems
are not appropriate for this application) has grown rapidly, driven by
the consumer market, in particular the photographic and cinematics
industries.  Many commercial applications have also become feasible,
including within agriculture, the film industry, search and rescue,
sports, and infrastructure inspection.  This has led to rapid
improvements in \uav technology and capability.

Table~\ref{tab:uavoverview} provides a performance summary for a
number of current commercially available systems.  With some
exceptions, flight time is typically limited to 40~minutes, and
payloads of up to 10~kg.  We note that in these cases, flight time
relates to that achievable by a \uav equipped with standard equipment
(battery capacity, propeller type), and therefore should be seen as a
minimum achievable.  Advances in battery technology means that flight
times are likely to extend to one hour for high-end systems within the
next few years.  The use of hydrogen fuel cells has been shown to
extend flight times beyond two hours, and commercial options are
becoming more readily available.  It should also be noted that
\uavs have flown at Mount Everest base camp, at an altitude
of about 5,300~m, showing that the thinner atmosphere at these heights
does not preclude \uav flight.  Optimisation of propeller designs can
also further improve performance at higher altitudes.

\begin{table}
  \begin{tabular}{llll}
    Model & Type & Payload & Flight time \\ \hline
    Yuneec Tornado& Hexa & 2~kg & 24 min \\
    DJI Matrice 600 & Hexa & 6~kg & 38 min\\
    DJI Agras & Hexa &10~kg & 24 min \\
    DJI S1000 & Octo &5~kg & 15 min\\
    Multirotor Eagle &Octo & 2.5~kg & 20 min \\
    Multirotor Skycrane & Octo &6.5~kg & 12 min\\
    Firefly&Hexa & 6.8~kg & 15 min\\
    Quarternium Hybrix 2 &Quad & 2kg & 120 min\\
    (petrol)\\
    HyDrone 1550 & Hexa & 5kg & 150 min\\
    (hydrogen fuel cell)\\
    
  \end{tabular}
  \caption{A performance summary for a number of commercial UAV
    systems.  Flight times are when hovering, and would not include
    ascent and descent times.  Current UAVs can ascend to 1~km in 2--5
  minutes, depending on payload and model, while descent times are
  usually slightly longer due to downdraught effects, i.e.\ the UAV
  descending into the turbulence that it has created.}
  \label{tab:uavoverview}
\end{table}

We present techniques to obtain instantaneous relative \uav position and some conceptual
designs for these \ags systems in \S2, along with preliminary
investigations that we have performed in \S3, including Monte-Carlo
\ao system modelling.  Our conclusions are
formed in \S4.

\section{Artificial guide stars using a UAV platform}

The concept that we describe here has several key requirements.  The
\uav would be required to maintain its position for the duration of a
science observation, or the ability to pause science integration while
\uavs are changed.  The \uav must be able to maintain a knowledge of
its relative lateral position to high precision (of order 1~mm).  Its
altitude must also be known, though with a somewhat relaxed precision,
such that measurement from the ground is sufficient.  The \uav
must be able to follow a pre-programmed flight path, and accept minor
adjustments to position while in flight.  The \uav must also carry a
payload with light sources, position sensors,
and a low latency communication link to the ground station, with
sufficient bandwidth to transfer sensor information (MBit/s) with
a latency of 1~ms or lower (below the atmospheric coherence time).

In this section, we discuss requirements and identify techniques that
can be used to improve \ao capability using the \ags concept.

\subsection{UAV-determined position}
\label{sect:uavpos}
There are a number of techniques that have the potential to determine
\uav instantaneous relative position with the required accuracy.  For
improved accuracy these techniques could be used in combination.
Lateral position is key for this application, however most of these
techniques will also be able to determine vertical position with
greater precision than is required.

\subsubsection{Real-time kinematics}
The use of \rtk information \citep{rtk} is a standard approach by
which position relative to a base station can be measured with typical
accuracy of order one centimetre, though accuracy decreases with
distance from the base station.  The \rtk technique works by using the
phase of \gnss carrier signals and the base-station link, rather than
just the content of these signals.  Although this precision is not
sufficient to meet the requirements here (approximately 1~mm), future
improvements are possible, and it can serve as a baseline measurement
since it provides position relative to a fixed ground station.

\subsubsection{Accelerometer and gyroscope measurements}
Accelerometer measurements can be integrated (twice) to obtain
position information (given an initial starting position and
velocity).  By using a 3-axis accelerometer combined with gyroscopic measurements (to
ensure that the correct Cartesian reference frame is used during
integration, and to aid gravity subtraction), a 3-d position vector
can be obtained (relative to an initial position).  Unfortunately, these
measurements will contain noise, which when integrated will lead to large
position errors rapidly building up, as discussed in \S\ref{sect:accel}.  We
therefore propose a solution by which the short-time-period position is
obtained using accelerometer and gyroscope measurements, while the
long-time-period position is obtained using some other absolute
position technique, for
example \rtk measurements.  These measurements can be
combined in an optimal way using a Kalman filter
\citep{Kalman60contributionsto,kalmangps}.
This is very similar to the SuperBIT telescope pointing system
described by \citet{superbit}, where guide star measurements are read
out slowly and high frequency gyroscopic data are integrated
in between.  However, in this case, a Kalman filter is not explicitly used.

The Kalman filter is designed to combine information in the presence
of uncertainty, namely the errors associated with accelerometer and
\rtk sensor information.  The uncertainties in absolute position (from
the \rtk sensor) will be relatively large (of order 1~cm), while
accelerometer uncertainty will be much smaller.  Fortunately, this
concept does not rely on an absolute position, but rather, a position
measurement relative to some starting point.  Combining the change in
position derived from accelerometer measurements (which update rapidly
within millisecond timescales) with the direct measurements from the
\rtk (updating at a few Hz) using a Kalman filter will therefore
provide a far more accurate relative position knowledge than using
only one of these sensors alone.

\subsubsection{Ground base stations}
\begin{figure}
  \includegraphics[width=\linewidth]{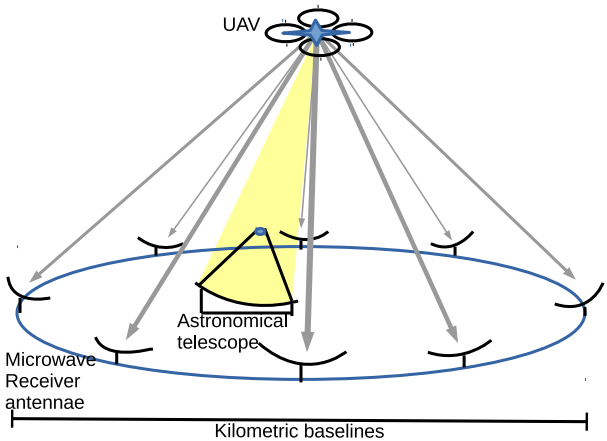}
  \caption{A figure demonstrating the use of multiple ground base
    stations to triangulate relative UAV position, using time of
    arrival measurements of a pulse emitted by the UAV.  The phase of
    a continuous signal could also be used.  In this figure, the grey
    arrows represent the microwave pulses, while the yellow beam is
    the AGS signal.  Microwave receiver telescopes are spread over a
    large area to improve triangulation accuracy.}
  \label{fig:basestations}
  \end{figure}

An alternative technique to estimate relative instantaneous \uav
position is the use of multiple ground base stations to receive a
microwave pulse or amplitude modulated continuous coherent signal from the
\uav.  The pulse time of arrival, or phase of a continuous signal at these
receiver stations can then be used to triangulate the \uav position,
relative to a calibrated intial location.
Atmospheric effects will lead to uncertainty in these measurements.
However, given a sufficient number of receivers, and by operating the
\uav at a height away from significant turbulence, we anticipate that
these uncertainties can be largely mitigated, leading to a
sufficiently accurate position estimation.
Fig.~\ref{fig:basestations} illustrates this technique.

The microwave pulse can be produced using a \race generator \citep{raceGenerator},
giving a well defined pulse shape and time.  Femto-second level timing
accuracy will be required to measure pulse arrival times with 1~mm
positional accuracy.

Since the complexity of these triangulation approaches is high,
requiring multiple ground stations and high precision timing, it will
not be cost effective compared with the on-board sensor approach
outlined previously.  We therefore do not consider it further here.

\subsubsection{Optical measurements}
We note here that the use of direct optical measurements to determine
\uav position, e.g.\ processing of images of the ground taken by the
\uav, does not provide a position estimation with sufficient accuracy
since these measurements will be affected by the atmospheric
turbulence that is to be corrected.  However, rotational information
can be obtained by direct optical measurement, supplementing
gyroscopic measurements.  A camera on-board the \uav, facing the
ground (or at night time, the sky), can be used to estimate the
instantaneous pitch, roll and yaw of the \uav, by correlating
image frames with a reference.

By combining the use of \rtk, accelerometer and gyroscope information
with imaging, and triangulation using multiple ground base stations,
improved position estimates can be obtained.  However, for simplicity,
we do not consider this further here, rather concentrating on the
accelerometer based approach.

\subsection{Multiple guide stars for tomographic tip-tilt
  determination}
\begin{figure}
  \includegraphics[width=\linewidth]{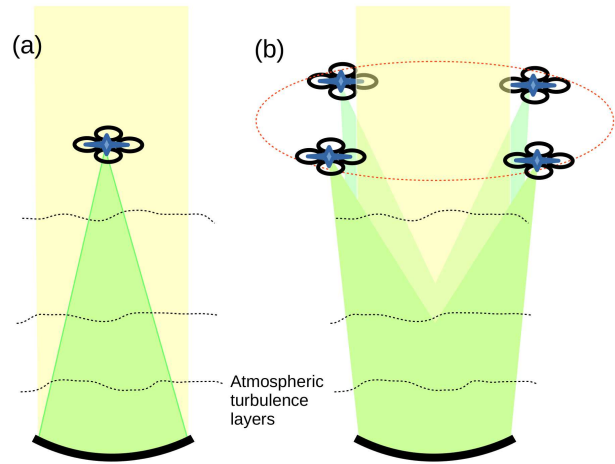}
  \caption{(a) Showing focal anisoplanatism due to the finite altitude
    of the UAV AGS light source.  (b) Mitigation of focal anisoplanatism
    using multiple AGS, which allow a larger volume of turbulence to
    be sampled.  Only turbulence within the green area will be
    measured by the wavefront sensor.}
  \label{fig:cone}
  \end{figure}

The \ags technique described here does not allow the full volume of
atmospheric turbulence to be probed, since the \uav is not at an
infinite distance from the telescope, as shown in Fig.~\ref{fig:cone}.  This therefore results an a rather extreme ``cone
effect'', or focal anisoplanatism \citep{1990A&A...235..549T}.  This is less
pronounced in the solar \ao case when observing at low zenith angles
(morning or evening) since the \uav can be further from the telescope,
i.e.\ it can be launched from a more suitable location (Fig.~\ref{fig:solar}).

A partial solution to this focal anisoplanatism is to use multiple
\ags and compute a tomographic wavefront reconstruction of the
turbulence
\citep{2004SPIE.5382..478E,2000Natur.403...25E,2002ESOC...58...11R,2002ASPC..266..562B,2001A&A...378..710T,1999A&A...342L..53R,ngslgstomoshort}.
This will allow the identification of turbulent wavefronts as a
function of height across a wider field of view, and therefore
projection along the telescope line of sight to obtain the integrated
wavefront in this direction.  The tip-tilt information can be obtained
with greater accuracy, and in particular, the tip-tilt signal
introduced by the strong atmospheric ground layer can be isolated.

For solar \ao, where high order wavefront sensor information is
required, the use of multiple \ags will yield improved \ao
performance, due to the reduction in focal anisoplanatism, and in
particular, improved determination of wavefront perturbations
introduced by the typically strong ground layer.

\subsection{Turbulence above the UAV}
\label{sect:lowlevelturb}
Since the \ags is at a finite altitude within the Earth's atmosphere,
there will usually be atmospheric turbulence above the \uav, which
will then be unsensed by the wavefront sensor.  This will therefore
degrade the \ao system performance.  This can be somewhat mitigated by
operating the \uav at a higher altitude, though physical restrictions
(limited flight times, ability to operate in the thinner atmosphere)
currently mean that the \uav cannot reach an arbitrary height.

Fortunately, at many astronomical sites, e.g.\ the Roque de los
Muchachos on La Palma, ground layer turbulence is responsible for the
majority of introduced wavefront phase perturbations.

During day time observations, \citet{2015JPhCS.595a2035T} find that
the vast majority of turbulence is at the ground, as high as 95\% in
some cases.

During night time observations, using the CANARY instrument on the
William Herschel Telescope at the Roque de los Muchachos observatory
on La Palma, \citet{martin2017} find that median seeing of the ground
layer (up to 1~km) is 0.59~arcsec, while the combined seeing of higher
altitudes is 0.21~arcsec with a standard deviation of 0.09~arcsec.

Likewise, \citet{2011MNRAS.416.2123G} find that between 67-76\% of
turbulence during median conditions is situated in the boundary layer,
with more than 95\% of this below 500~m, at this site.

At the Cerro Paranal site, \citet{2012MNRAS.420.2399M} find that for
median conditions, the atmosphere below 1~km contributes to a seeing
of 0.9~arcsec, while above this introduces about 0.45~arcsec.

At Teide observatory, \citet{2011MNRAS.410..934G} find that 60\% of
turbulence is within the boundary layer (up to 1~km) for the average
profile, increasing to 72\% for the median profile, with 85\% of
optical turbulence in the lower layers (below 5~km).

On Mauna Kea, \citet{2005PASP..117..395T} find that optical turbulence
in the first 700~m above ground contribute to typically half of the
total integrated seeing.  

It should also be noted that these seeing estimates generally do not
include dome seeing, which can be a large contribution to observed
seeing at a telescope (though varies depending on dome structure, site
and wind direction).  Therefore, the ground layer turbulence as seen
by an \ao system will usually constitute a greater fraction of overall
turbulence than is reported from external seeing monitors.  Therefore,
significant improvement in optical quality will be possible by just
correcting for these ground layers.  For the astronomical case, only
higher layer tip-tilt would remain uncorrected.  For the solar case,
higher order information above the \uav height would remain
uncorrected.

\subsection{Optical leverage of turbulence close to the UAV}
If the \uav light source is located close to a turbulent layer, then
the patch of this turbulence viewed by the telescope will be small due
to focal anisoplanatism.  A small tilted region of phase can therefore
impart a larger shift in observed source position
\citep{1997A&AS..121..569R}.  However, by operating the \uav at
heights away from strong turbulence layers, this effect can be largely
mitigated, particularly for smaller telescopes.  Increasing the
altitude of the \uav also reduces the strength of this effect, as
focal anisoplanatism is less severe.  The simulations which we present
in \S\ref{sect:sim} implicitly include this effect.

\subsubsection{UAV signals combined with LGS uplink tomography}
It has been shown that for tomographic \ao systems, some \lgs tip-tilt
information can be obtained for higher layer turbulence
\citep{reevesThesis}.  Therefore, by combining this information with
tomographic \uav tip-tilt information (which includes lower layer
atmospheric tip-tilt information), improved tip-tilt information can
be retrieved, and hence \ao correction improved.

\subsection{Science field obscuration}
Placing a \uav within a telescope's field of view has the potential to
block light and obstruct the point spread functions of the science
images.  Downdraught produced by the \uav may also introduce some
wavefront perturbations.  However this can largely be mitigated by
operating the \uav at an altitude away from inversion layers,
i.e. where the atmosphere is at a constant temperature.  In this case,
although the downdraught will be turbulent, changes in refractive
index (caused by temperature changes) will largely not be present.
The typical turbulent cell size produced by a \uav, and dynamics of
these, is currently under investigation.  Additionally, if the \uav is
kept downwind of the telescope, any \uav-produced turbulence will not
be seen in the science image.  Unlike a laser guide star, a waveband
filter cannot be used because the \uav is physically present at the
guide star location, i.e.\ the \uav will block some light to the
science detector, and modify the instrumental \psf.  This will
obviously have an impact on the astronomical science, and therefore
should be minimised for science cases where accurate \psf shape is
important.

\begin{figure}
  \includegraphics[width=\linewidth]{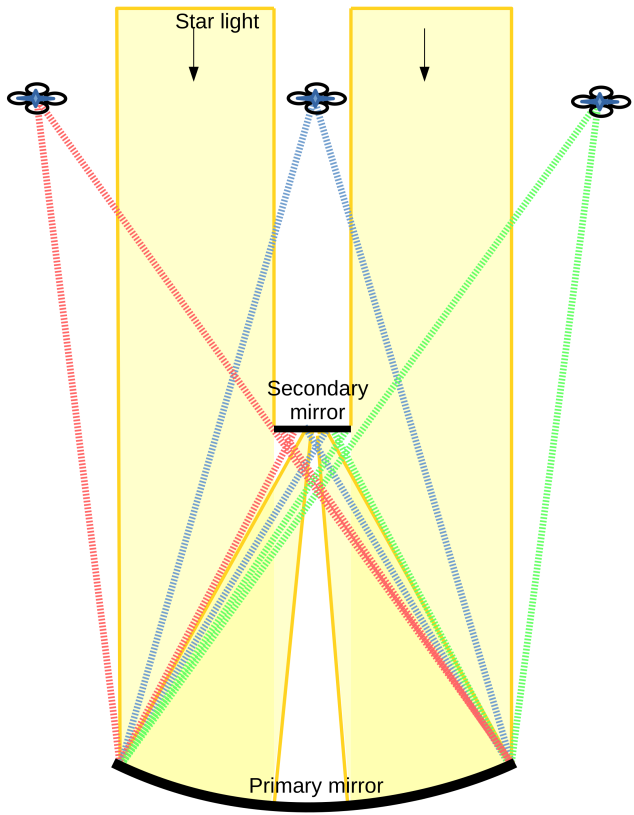}
  \caption{A figure showing UAV locations that would not interfere
    with science measurements, behind the central obscuration (with
    the AGS light path shown in blue), or
    beyond the field of view (with the AGS light path shown in green
    and red).  Star light is represented by the column of yellow,
    coming from a source at infinity.}
  \label{fig:obscuration}

\end{figure}

There are several possible solutions.  First is the
possibility of stationing the \uav behind the telescope central
obscuration.  Since the wavefront sensor for the \ags is not focused
at infinity, it will therefore see the \uav, while the science path
will not be affected.  This is particularly appropriate for larger
telescopes, where the central obscuration is larger, as shown in
Fig.~\ref{fig:obscuration}.  We note that as the telescope field of view or
\uav height increases, the area over which the \uav can remain hidden
decreases for a non-zero science field of view.

The second possibility is to station multiple \uavs outside the
telescope scientific field of view, and perform a tomographic
wavefront reconstruction.  This introduces additional complexity, but
has the advantage that the focal anisoplanatism can be mitigated (as
discussed previously), and performance across the field of view
increased.

In both of these cases, care must be taken to ensure that the \uav
does not drift into the science field (e.g.\ due to a gust of wind).
It is possible that mitigation steps can be taken, i.e.\ if the \uav
moves too far from its expected position, the light source could be
turned off to avoid polluting science images (though this would result
in the loss of tip-tilt information, particularly if only one \uav is
in use), or a narrow band filter could be used on the science path to
block the \uav signal (though obscuration would still occur).

Stray light from the \uav source due to multiple Rayleigh scattering
may also be present, and so it may be desirable to have a narrow band
filter present to stop this.  However, if the \uav is faint this is
unlikely to be a problem: since the \uav used for a tip-tilt signal
only (not higher orders) on a large telescope aperture, a signal of
15--16~th Magnitude may be usable, and therefore Rayleigh scattering
will be negligible.

For an 8~m telescope with a \uav at 1~km at the edge of the on-axis
science field (i.e.\ the cylindrical observation beam) the \uav would
appear about 13~arcmin off axis.  This is a large field of view for an
astronomical \ao system, and it is highly likely that a suitable \ngs
could be found within this field, particularly within the galactic
plane, meaning that sky coverage would approach 100\%: the \uav would
not be necessary.  However, for a \uav at 4~km, this field narrows to
3.5~arcmin, and at 10~km, to 1.4~arcmin, at which point, \ngs sky
coverage becomes more severely restricted, particularly outside the
galactic plane.  For a 4~m telescope, a 10~km \uav could be placed
about 40~arcsec off-axis without obscuring the on-axis science field.
Therefore, the higher the \uav can operate, the larger the gain in
achievable sky coverage, though this does not affect an on-axis \uav
behind the telescope central obscuration.  Here, we have considered
only the geometrical shadow of the \uav in the optical beam, rather
than the Fresnel effect.  However, if the light source was suspended
at the side of the \uav, rather than centrally mounted, the Fresnel
effect would be small.

\subsection{UAV flight time limitations}
Flight time of current \uavs is currently restricted to less than one
hour by limitations in battery technology.  Hydrogen fuel cell
developments have shown that flight time can be increased to more than 2
hours.  However, even though battery and fuel cell technology is
rapidly improving, short flight times will still be the reality
(compared with night-long science observations), particularly since time
required to reach altitude and return safely must be included.  This can
therefore be problematic when long astronomical science exposures are
required, and there are two possible solutions that we have
identified.

\subsubsection{In-observation swapping of UAVs}
The first solution is to perform in-flight swapping of \uavs, such
that a fully charged replacement is scheduled to take the position of
a \uav that needs to return for recharging.  During this operation,
the new \uav would manoeuvre to the required altitude, just beyond the
telescope field of view. This operation would occur in between science
integrations, which are typically split into individual exposure of
not more than 30~minutes each.  The \ao loop would be disengaged, and the \uavs would swap position.  The new \uav
would then be acquired by the wavefront sensor and the \ao loop engaged, and once
stabilised, the next science exposure started.  With full automation, this
procedure could be expected to have negligible impact on most science
observations, and would be expected to result in less than one minute
of down-time.

\subsubsection{In flight recharging}
A second solution for limited flight time is to develop in-flight
recharging capabilities, based on wireless energy transmission
\citep{uavRecharging}.  This has previously been demonstrated with a
helicopter using microwave energy transfer \citep{helicopterRecharging}.  This would require a ground-based
high power microwave transmitter, which would send a directed beam to
a receiver on the UAV, where the energy would be collected and
stored.  We note however that we have not investigated the practicality of this scheme.

\subsection{Autonomous operation}
The \uav application proposed here would operate autonomously, to
remove the possibility of pilot error.  A fully automated system would
enable the \uavs to depart from the base station, manoeuvre to the
correct position on-sky ahead of telescope acquisition, and then
automatically follow the course of the position change due to
telescope tracking.  Position offsets derived from the wavefront
sensor signals by the \ao system can be sent to the \uav to perform a
slow guiding offload.

At the end of operation, the \uav would return to the base station,
and automatically begin recharging.  Therefore once such a system has
been commissioned, minimal operator intervention would be required beyond
routine maintenance.

\subsection{In-flight safety}

\subsubsection{UAV component failure risk}
The largest contribution to risk associated with \uav operation comes
from pilot error, which is largely mitigated by having a fully
autonomous system.  However, the introduction of redundancy and
fail-safe operation should also be considered.  Should a \uav component
fail during operation (e.g.\ a motor), it must be possible for the
\uav to return safely to the ground.  For this reason, we recommend
the use of hexa and octocopters, which are able to fly and land safely
in the event of motor failure.

Communication failure is also a possibility, in which case, the \uav
should be programmed to return automatically to its launch location,
based on a \gnss signal.

Battery failure should be considered, and an auxiliary power supply
included. 

\subsubsection{Collision risk}
The risk of a collision of a \uav with the telescope should be
considered.  However, since telescopes do not usually operate at
zenith, a failed \uav is unlikely to fall onto the telescope, though
other telescopes on-site might be affected.  The
\uav would be expected to carry a parachute and a fail-safe mechanism
to stop the rotors, to lessen impact velocity and reduce collateral
damage to people, other buildings and infrastructure (and the \uav
itself).  

When multiple \uavs are in operation, there is the risk of in-flight
collision.  However, this will be largely mitigated by having all
flight under computer control, with algorithms to explicitly avoid
this scenario, in addition to onboard proximity sensors.  Collisions
with other aircraft is also possible.  The airspace above some
observatories is a designated no-fly zone, and therefore these
observatories could operate a \uav without collision risk.  At other
observatories, aircraft spotter systems would be required, as is
already the case for \lgs systems.  A mode S transponder would be
carried by all \uavs so that they are uniquely identifiable.  The
\uavs would also carry proximity sensors to aid collision avoidance
with rogue \uav operators.  If a collision with a larger aircraft was
predicted, it would be the responsibility of the \uav to take
avoidance action.  Collisions with birds should also be considered.

\section{Modelling investigations}
\label{sect:sim}
Key to the success of this \uav-aided \ao technique is the ability to
derive the instantaneous relative position from accelerometer
measurements.  Unfortunately, the double integral required to obtain the
position from acceleration data can lead to a rapid (1.5 power) build up of
error.  We therefore investigate the necessary accelerometer
specification required to meet the position knowledge accuracy.  Here,
we consider only accelerometer measurements, and do not consider the
integration with absolute position estimates (\rtk), since it has
previously been shown that a Kalman filter can be used to optimally
combine these measurements.  Rather, we concentrate on determining the
accelerometer specifications (polling rate and uncertainty) that would
be required to meet position estimates for a given period of time.

\subsection{Accelerometer position accuracy}
\label{sect:accel}
We have investigated the accelerometer parameters needed to maintain a
position knowledge to within a specified error over a specified time
period.  In our modelling we include a random thermal (Gaussian) noise
uncertainty on the accelerometer output, and also the effect of
digitisation of the accelerometer signal.  For simplicity, we consider
a one dimensional system, and include cases where the accelerometer is
at rest, moving with a sinusoidal motion, and undergoing constant
acceleration.

\subsubsection{Accelerometer at rest or under constant acceleration}
We first consider the case of an accelerometer at rest, and input
different noise levels into the accelerometer readout process.  Using
a readout (sampling) rate of 100~Hz (time period of 10~ms), we measure the
accelerometer-predicted position over 100~s.  We repeat the
measurements 100 times.

\begin{figure}
  \includegraphics[width=\linewidth]{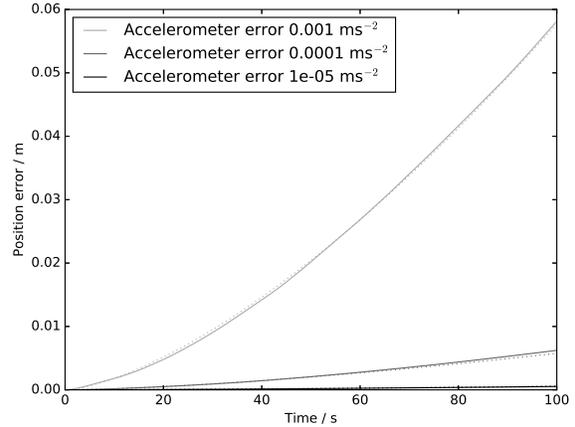}
  \caption{Accelerometer-derived position error as a function of time
    for different accelerometer noise levels as given in the legend,
    in ms$^{-2}$.  Solid curves are Monte-Carlo measurements, while
    dotted curves show Eq.~\ref{eq:pos}.}
  \label{fig:accelModel}
  \end{figure}

Fig.~\ref{fig:accelModel} shows how the position error grows as a
function of time for different accelerometer noise levels, given in
ms$^{-2}$.  When the noise is low, accelerometer position remains
accurate to less than 1~mm for 100~seconds.  However, in all cases,
the position error grows faster than linearly.  \citet{THONG200473}
show that position error due to accelerometer uncertainty grows as
\begin{equation}
  ERR(t) = \frac{1}{\sqrt{3}}\frac{\sigma}{\sqrt{f}}t^{1.5}
    \label{eq:pos}
\end{equation}
where $t$ is the time, $\sigma$ is the rms accelerometer uncertainty
and $f$ is the sampling frequency.  We find that our modelling is in
good agreement with this.

\begin{figure}
  \includegraphics[width=\linewidth]{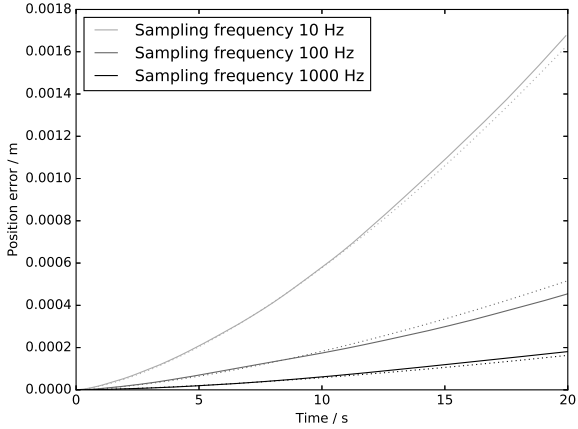}
  \caption{Accelerometer-derived position error as a function of time
    for different accelerometer sampling rates as given in the legend,
    in ms$^{-2}$.  Solid curves are Monte-Carlo measurements, while
    dotted curves show Eq.~\ref{eq:pos}.}
  \label{fig:accelModelDT20}
  \end{figure}

Fig.~\ref{fig:accelModelDT20} shows how the sampling period affects
position error.  Here, we use an accelerometer uncertainty of
0.0001~ms$^{-2}$ at every sample point.  Easily obtainable commercial
accelerometers offer sampling rates above 300~Hz, with error
(uncertainty) lower than 0.0001~ms$^{-2}$ \citep{innalabsWhitepaper}.
We can therefore see that accelerometer measurements will be able to
estimate position with the required accuracy for significant periods
of time, beyond which the 1.5 power  growth in error can be constrained
by \rtk measurements using a Kalman filter.

\begin{figure}
  \includegraphics[width=\linewidth]{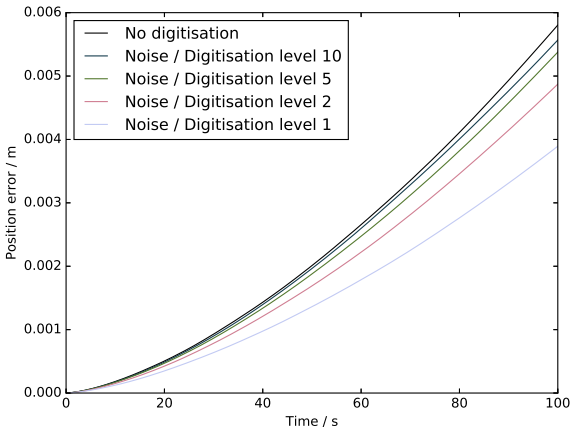}
  \caption{Accelerometer-derived position error as a function of time
    for different accelerometer digitisation levels as given in the legend,
    in ms$^{-2}$.  The value of digitisation given here is the number
    of times larger than the digitisation level one standard deviation
  of noise is.  Therefore, coarser digitisation is given by a lower
  number.  In this case, sampling rate is 100~Hz, and accelerometer
  noise is 0.0001~ms$^{-2}$.}
  \label{fig:accelModelDig}
  \end{figure}

Digitisation (quantisation) of accelerometer signals will occur when
the analogue current or voltage generated by the accelerometer is
converted to a digital form.  If the accelerometer is stationary (or
moving with no acceleration), then it can be expected that
digitisation of the signal will actually improve performance: if
digitisation is taken to the extreme, then all measurements would be
exactly zero, and thus, the derived position would also remain
zero. This is verified in Fig.~\ref{fig:accelModelDig}.  However,
since the purpose of an accelerometer is to detect changes in motion,
study of a static accelerometer is an academic exercise only.

\subsubsection{Accelerometer with sinusoidal motion}
\begin{figure}
  \includegraphics[width=\linewidth]{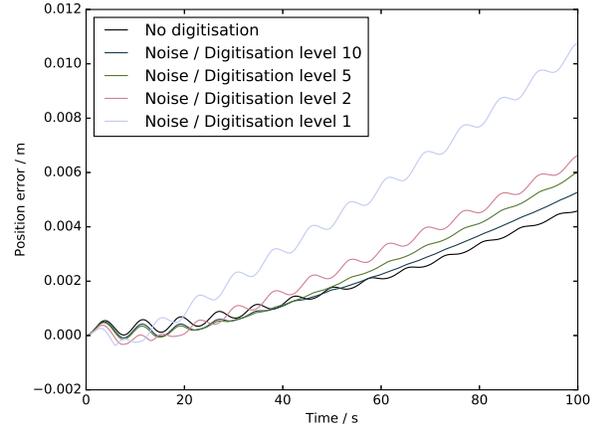}
  \caption{Accelerometer-derived position error as a function of time
    for different accelerometer digitisation levels as given in the legend,
    in ms$^{-2}$.  The value of digitisation given here is the number
    of times larger than the digitisation level one standard deviation
  of noise is.  Therefore, coarser digitisation is given by a lower
  number.  In this case, sampling rate is 100~Hz, and accelerometer
  noise is 0.0001~ms$^{-2}$.  The accelerometer is undergoing
  sinusoidal motion with a 10~cm amplitude and 7~s period.}
  \label{fig:accelModelSineAvDig}
  \end{figure}

Since the \uav will not be stationary, it is important to model an
accelerometer-derived position with non-uniform acceleration.  To this
end, we consider an accelerometer undergoing sinusoidal motion with
10~cm amplitude and 7~s period.  This amplitude is chosen as
representative (though pessimistic) of \uav stability in conditions of
light wind.  Accelerometers do not measure instantaneous acceleration,
but rather that integrated over a period of time.  We therefore
include this in our model, using a high resolution time step to
generate actual position, while integrating the acceleration over this
period.  Fig.~\ref{fig:accelModelSineAvDig} shows position error as a
function of time for a 100~Hz accelerometer sampling rate, with
0.0001~ms$^{-2}$ noise.  Different digitisation levels are shown, and
it is clear here that coarser digitisation gives worse performance.
However, the accelerometer-derived relative position estimate is
accurate to within 1~mm for about 40~s, which again, is sufficient.

\ignore{
\subsection{Kalman filtering}
In a real system, in addition to the accelerometer-derived position
measurements, actual position can be measured using \gnss and \rtk
systems.  However, the position error is much larger.  Therefore, by
combining these two sources of information using a Kalman filter,
improved position knowledge can be achieved.  Essentially, the
accelerometer is used to measure precise position changes at a high
rate, while the \rtk information is used with a low pass filter to
avoid the rapid increase in accelerometer measurement position error
that can otherwise arise.
}

\subsection{Monte-carlo modelling of AO performance}
To investigate the \ao system performance gain that a \uav \ags can
deliver, we use \thedasp \citep{basden5}.  This Monte-Carlo modelling software has been widely used for \ao
system modelling \citep{2013RAA....13..875J,basden8,dicure,basden15,2010aoel.confE8003Mshort,basden17,2015MNRAS.450...38J,basden21}, and
includes an end-to-end model of the atmosphere, telescope and \ao system.

\subsubsection{Model parameters}
We investigate the performance of a \ltao system on a 8~m telescope,
using 4 \lgss.  No \ngss are used, and the \lgs tip-tilt signal is
assumed invalid.  We assume a sodium layer centred at 90~km with a
10~km \fwhm, and explore two different \lgs asterisms, with diameter
of 20~arcsec and 60~arcsec (the 4 \lgss are equally spaced on a circle
with this diameter).  We use
$16\times16$ sub-apertures for each \lgs \wfs, and assume the use of
low noise \emccds and high photon return from the sodium layer.  
The height of the \uav is investigated, and we use a single
sub-aperture Shack-Hartmann sensor for measurement of the \uav
tip-tilt signal.  Each \wfs operates at 589~nm.  The \dm has
$17\times17$ actuators, controlled using the \lgs signals, and a
separate tip-tilt mirror is used, controlled using the \uav \ags signal.

We use a standard \eso 35 layer atmospheric profile \citep{35layer},
with a Fried's parameter of 15.7~cm and a 30~m outer scale.  The
\ao system performance is measured on-axis at H-band (1650~nm).  We
use an \ao system update rate of 250~Hz, and integrate for 20~s, by
which time the \ao corrected \psf is well averaged.  A 1.2~m
telescope central obscuration is assumed.  The \ao system loop gain is set
to 0.5.

The \uav is modelled as a point source.  We first assume perfect
position knowledge, and then also model Gaussian uncertainty in the
measured position of the \uav with rms position uncertainties of 0.5,
1 and 2~mm.  In this case, at each simulation time-step, a Gaussian
random variable is added to the \ags tip-tilt measurements with a
standard deviation equal to the uncertainty in arcseconds
(i.e. approximately the arctangent of the position uncertainty divided
by \uav height).  We use this position uncertainty to model the
imperfect position measurement system on the \uav.  We note that
assuming a Gaussian distribution for position uncertainty is
pessimistic: this error will grow non-linearly with time between \rtk
measurements, and therefore is not well approximated by a Gaussian
distribution.

\subsubsection{Modelling results}
\begin{figure}
  (a)
  \includegraphics[width=\linewidth]{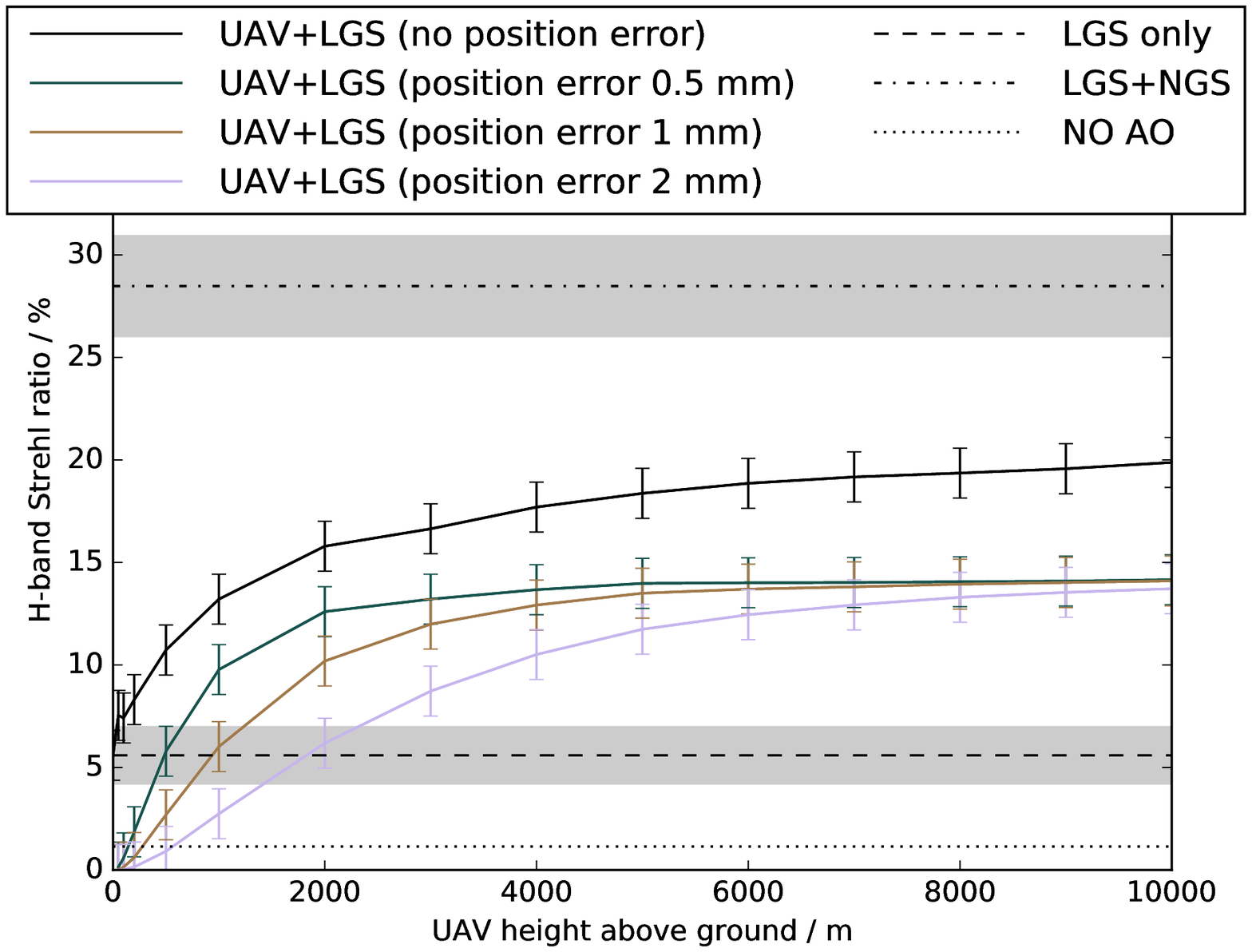}
  (b)
  \includegraphics[width=\linewidth]{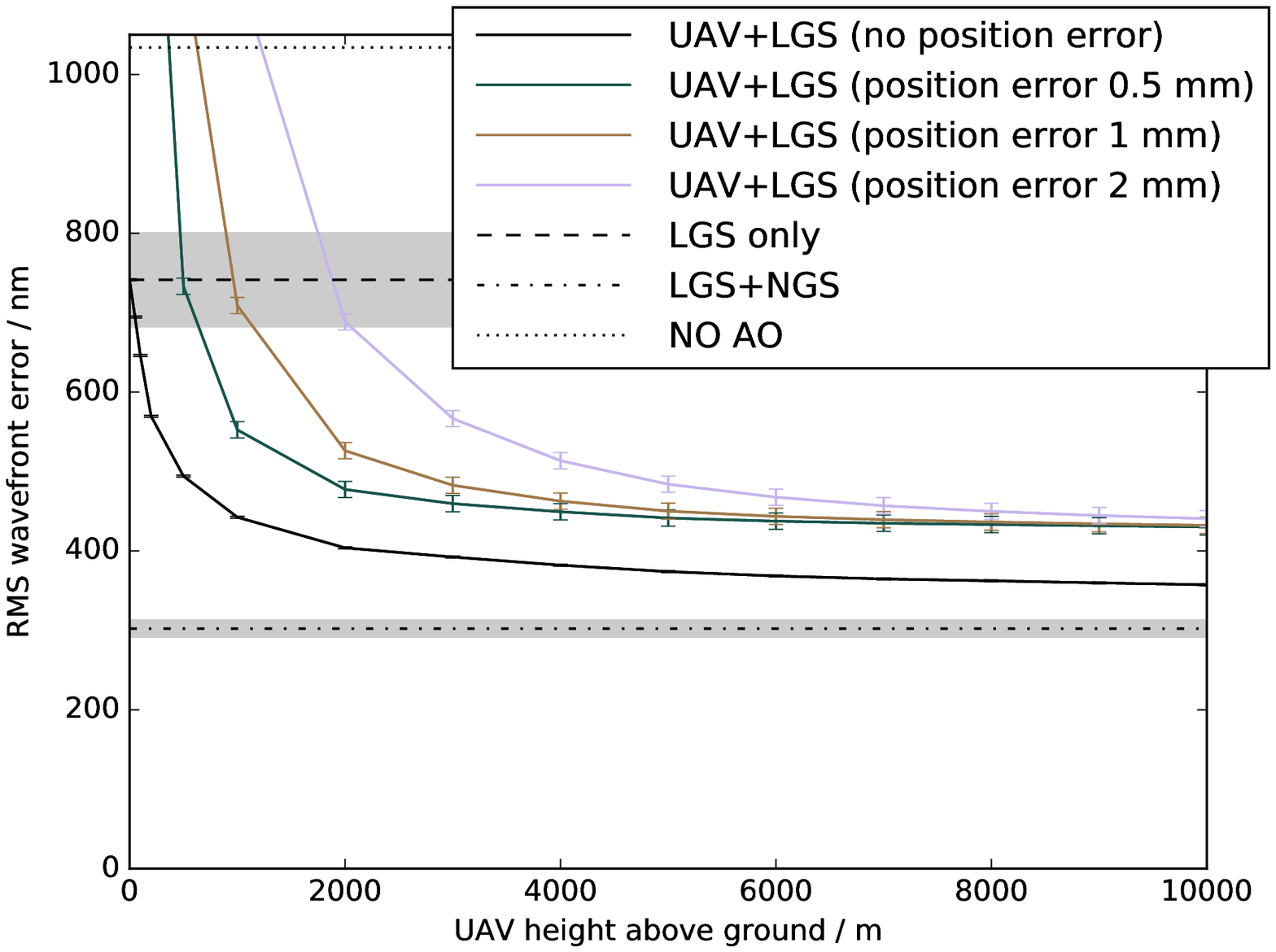}
  \caption{(a) AO system performance (H-band Strehl) as a function of UAV
    height.  The uncertainty in UAV position is given in the legend.
    A comparison with no AO, LGS-only AO, and full NGS+LGS AO is also
    shown.  The LGS asterism diameter is 20~arcsec.  The grey regions
    show the measurement uncertainty.  (b) As for (a), but showing
    residual wavefront error instead of Strehl ratio.}
  \label{fig:aoperf}
  \end{figure}

Fig.~\ref{fig:aoperf} shows the increase in H-band Strehl as a
function of \uav altitude, along with performance without \ao,
performance using only \lgss (i.e.\ no tip-tilt correction), and
performance using full \lgs and \ngs \ao (with an on-axis \ngs
tip-tilt sensor instead of a \uav, assumed to be bright,
i.e.\ negligible noise, with the same pixel scale as the \uav tip-tilt
sensor).  This clearly shows that the use of a \uav \ags can have a
significant \ao performance benefit.  Increasing \uav altitude is
beneficial, but even at 1~km, the improvement in \ao performance can
be significant if position uncertainty is small, compared to the use
of \lgss alone.  We see that at high altitudes, when accelerometer
error is low, performance levels are significantly better than those
without the \uav, though remain slightly short of the \lgs and \ngs
combined case, dominated by focal anisoplanatism.

It is evident, as would be expected, that the uncertainty in the \uav
position estimate can have a significant impact on performance.  In
the particular case that we model here, if \uav position uncertainty
is 1~mm, then the \uav must be at an altitude greater than 1~km before
\ao performance improvement is seen.  If the position uncertainty is
reduced to 0.5~mm, the \uav can operate at a lower height (above
500~m) to achieve some performance benefit.  It should be noted that
there is always a benefit in operating the \uav at a greater altitude,
regardless of the position uncertainty level.  We note that these
results will be dependent on atmospheric parameters, and that a higher
altitude is always desirable.

\begin{figure}
  \includegraphics[width=\linewidth]{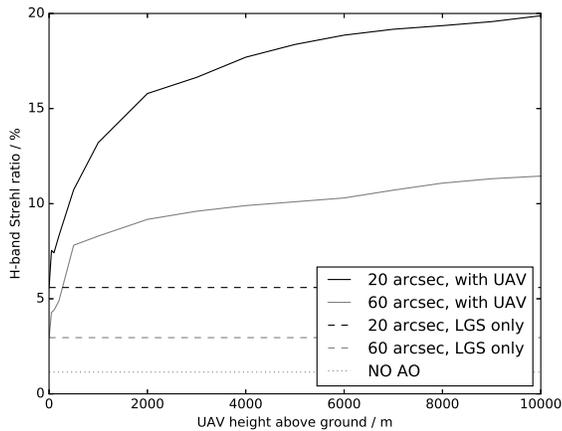}
  \caption{AO system performance (H-band Strehl) as a function of UAV
    height, for a LGS asterism diameter of 20 and 60~arcsec, as given
    in the legend.  Relative UAV position estimation is assumed to be accurate.}
  \label{fig:ast}
  \end{figure}

Fig.~\ref{fig:ast} compares \ao performance with two different \lgs
asterism diameters, and it is evident that a single \uav \ags is able
to improve \ao performance in both cases.

We note that the modelling presented here should not be taken as an
absolute performance reference, since this is highly dependent on the
atmospheric profile and turbulence strength, which vary from night to
night, and from observatory to observatory.  Rather, these results
should be taken as showing that the concept is valid, and that a \uav
\ags can be used to improve \lgs-only \ao system performance.

\subsubsection{Impact of outer scale}
\begin{figure}
  (a)
  \includegraphics[width=\linewidth]{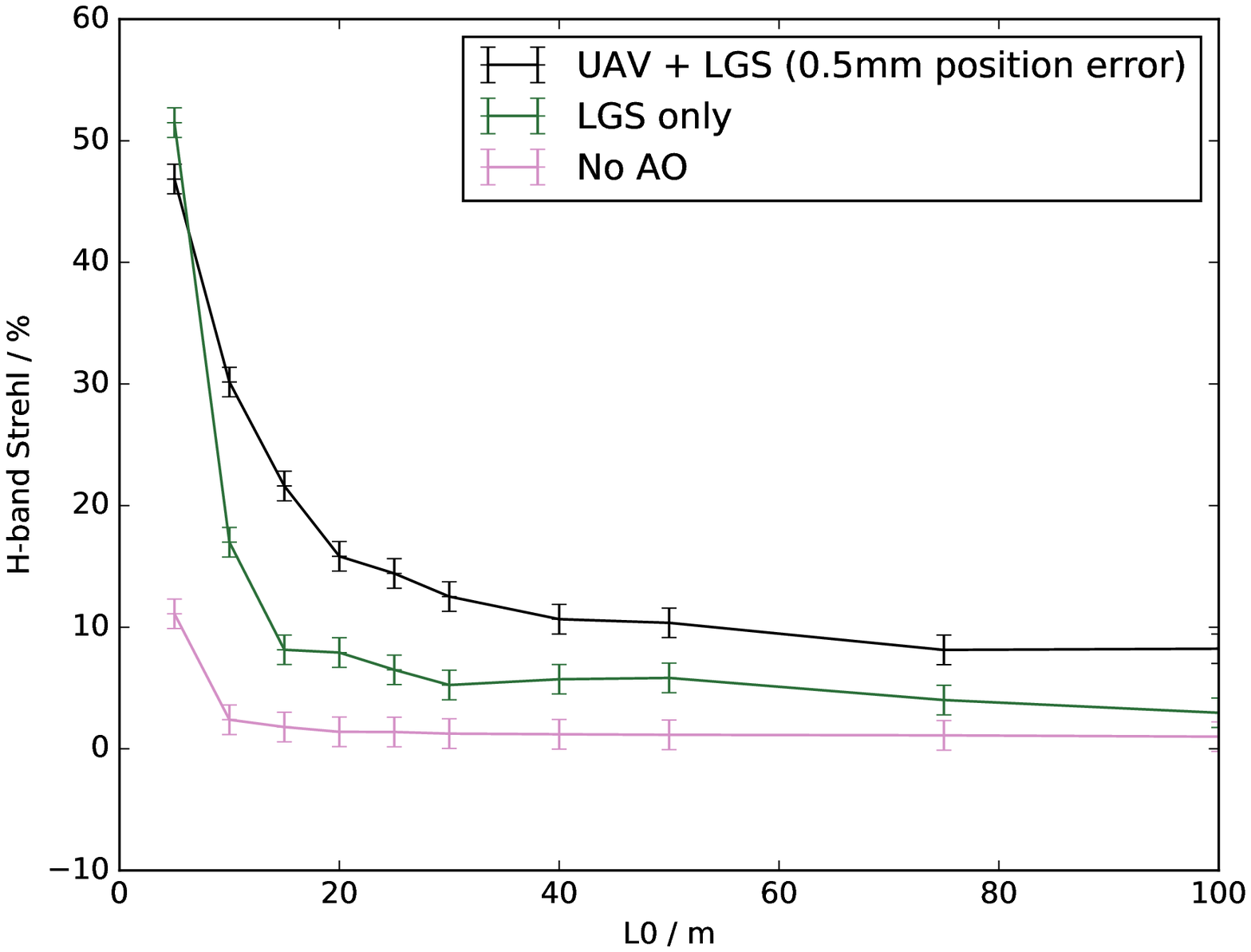}
  (b)
  \includegraphics[width=\linewidth]{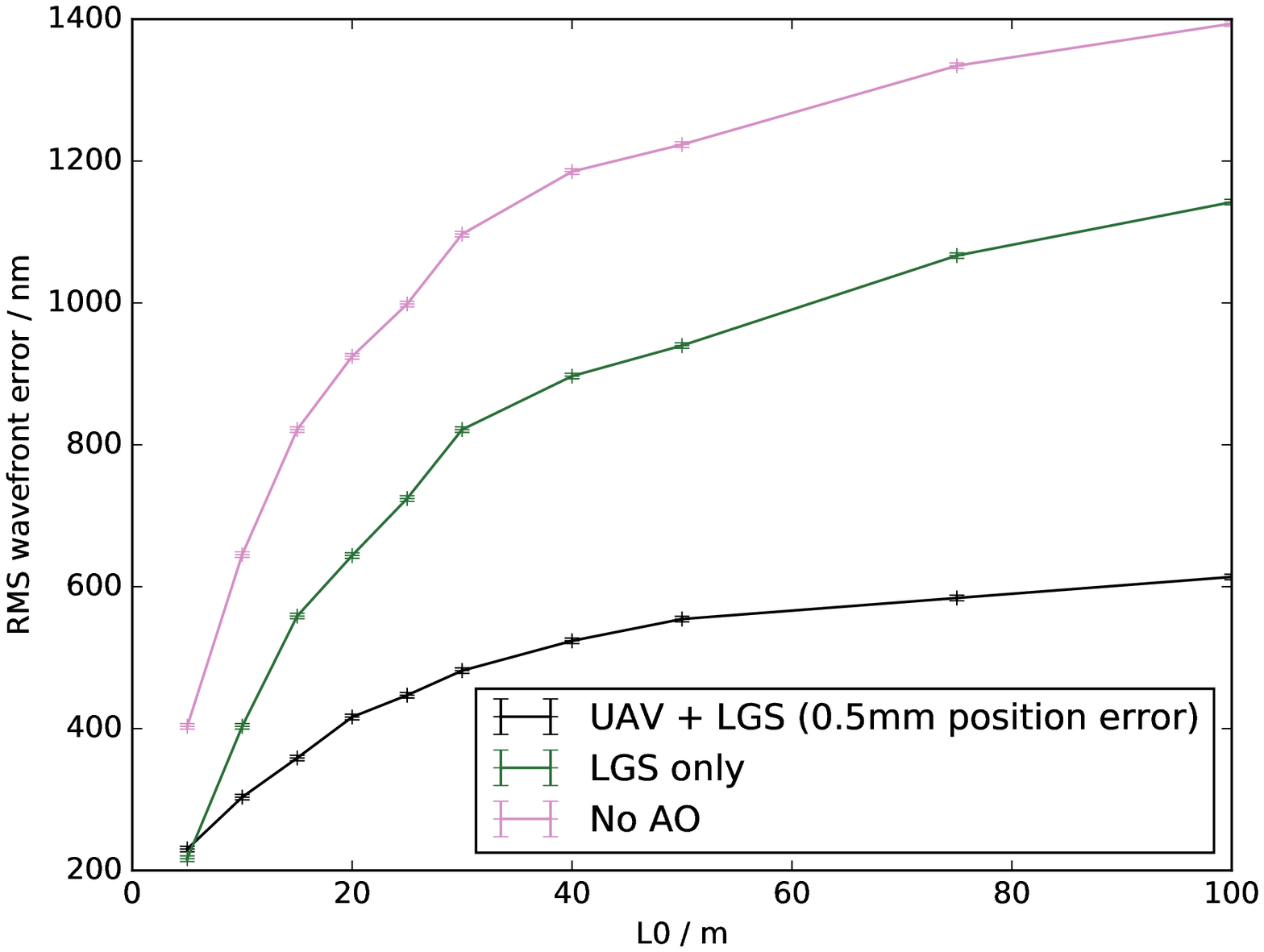}
  \caption{(a) A figure showing AO performance (H-band Strehl) as a
    function of atmospheric outer scale, for a \uav at 2~km with
    0.5~mm position uncertainty for the UAV.  (b) As for (a), but showing residual
  wavefront error instead of Strehl ratio.}
  \label{fig:outerscale}
\end{figure}

Within the literature, there is much debate about the range of values
for the atmospheric outer scale.  In our modelling we have selected
30~m as the default value.  However, we also investigate the \ao
performance as a function of outer scale from 5--100~m, as shown in
Fig.~\ref{fig:outerscale}.  Here, it can be seen that for a \uav at
2~km, performance is always improved, regardless of the outer scale.

\subsubsection{Physical propagation models}
\label{sect:fresnel}
\begin{figure}
  (a)\\
  \includegraphics[width=\linewidth]{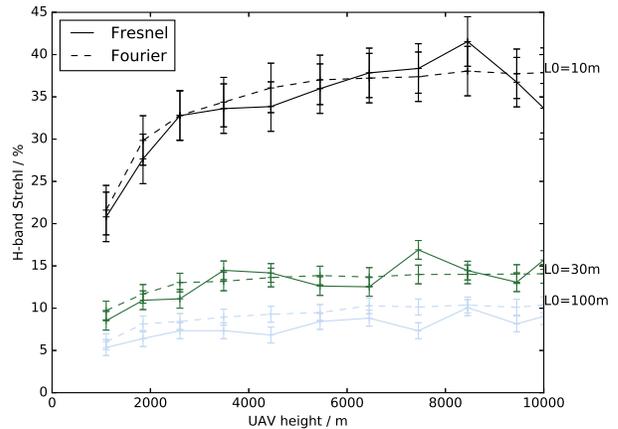}
  (b)\\
  \includegraphics[width=\linewidth]{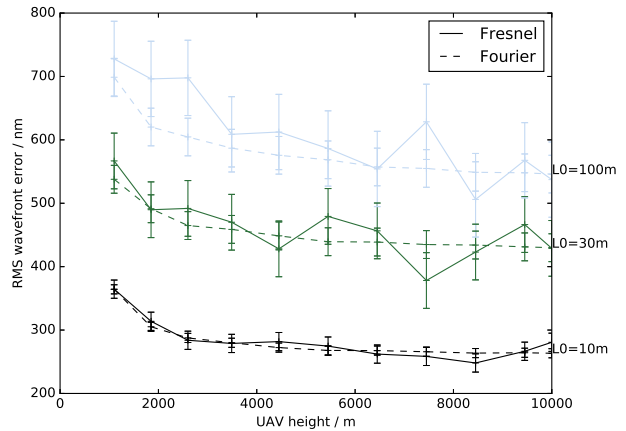}
  \caption{(a) AO system performance (H-band Strehl) as a function of UAV
    height for different atmospheric outer scale values, using
    Fresnel propagation models.  (b) As for (a), but showing residual
    wavefront error instead of Strehl ratio.}
  \label{fig:fresnel}
\end{figure}

The Monte-Carlo results presented within this paper use Fourier
propagation of the optical wavefront, i.e.\ converting between
far-field (pupil plane) to focal plane.  However, we also investigate
the use of Fresnel propagation for the \ags source, since this source
can be close to the perturbing layers.  Fig.~\ref{fig:fresnel} shows
that the difference between Fresnel and Fourier propagation results is
small.  At the largest outer scales, Fresnel propagation seems
to be slightly pessimistic compared with Fourier propagation, though
the difference is typically small (25~nm rms error).  We have
therefore used Fourier propagation for the rest of this paper, due to
reduced computational complexity.

\subsection{Future work}
This \uav \ags concept is in its infancy, and therefore much future
work is required.  Laboratory demonstration of \rtk and accelerometer
position estimation are under investigation.  \uav characterisation is
also required, and we are not assuming that the \uav is a ready-made
platform, but rather considering how the stability and aerodynamic
aspects of the \uav will affect performance.  Investigation of \uav
turbulence is also being carried out.  These investigations will be
followed by on-sky testing and verification of this concept, in
particular of the ability to provide an \ags reference.

We have also not measured the performance gain achievable with this
\uav technique when applied to \elt-scale telescopes.  Here, due to
the larger telescope diameter, more precise position knowledge is
likely to be required for the same gain in Strehl ratio (though other
performance metrics such as ensquared energy may perform better): the
telescope diffraction limit scales inversely proportional to telescope
diameter.

\section{Conclusions}
We have presented a novel concept using \uavs to provide an \ags
signal for adaptive optics systems, allowing full sky coverage to be
achieved for \ao corrected observations.  For astronomical \ao, this
concept uses a \uav system to provide tip-tilt signals, with higher
order correction being performed using \lgss.  This enables full sky
coverage, as \ngss are no longer required.  For solar \ao, this
concept uses the \uav \ags to provide a high order \wfs signal for
solar limb observations, where the solar structure itself cannot be
used for \ao correction.  We propose a system using \rtk,
accelerometers and gyroscopes to provide an accurate instantaneous
\uav relative position estimate.  Modelling for 8~m class telescopes
shows that the \uav should operate in excess of 1~km (with higher
altitudes being more favourable), and we note that this is within the
range of current \uav technology.  We find that Strehl ratio can be
increased by a factor greater than two compared to the case of
\lgs-only \ao for the cases studied here.  As \uav altitude increases,
\ao performance improves, and a \uav at 10~km is able to mitigate the
vast majority of wavefront tip-tilt, though performance does not quite
reach that achieved using a \ngs.

\section*{Acknowledgements}
This work is funded by the UK Science and Technology Facilities Council
grant ST/K003569/1, consolidated grant ST/L00075X/1 and an Impact
Acceleration Award. RM is supported by the Royal Society.  We also
thank an anonymous referee for their comments which have greatly
improved this manuscript.

\bibliographystyle{mn2e}

\bibliography{mybib}
\bsp

\end{document}